# Transparent Metamaterial Absorber with Broadband RCS Reduction for Solar Arrays

Xiangkun Kong[1,2*], Shunliu Jiang[1], Lingqi Kong[1], Qi Wang[1], Haobin Hu[3], Xiang Zhang[3], Xing Zhao[1]

[1]College of Electronic and Information Engineering, Nanjing University of Aeronautics and Astronautics, Nanjing, China
[2]State Key Laboratory of Millimeter Waves, Southeast University, Nanjing, China
[3]School of Mechanical Engineering, Nanjing University of Science and Technology, Nanjing, China

Correspondence should be addressed to Xiangkun Kong：xkkong@nuaa.edu.cn

Solar arrays are the primary energy source of the satellite. In this paper, a metamaterial absorber for solar arrays with simultaneous high optical transparency and broadband microwave absorption is presented. By tailoring the reflection response of meta-atoms, 85% absorption performance from 6.8GHz to 18GHz is obtained. In the meantime, by employing transparent substrates, including indium tin oxide (ITO) film and anti-reflection glass, a maximum of 87% light transmittance is achieved. The absorptivity of the proposed metamaterial absorber is simulated and measured experimentally. Light transmittance and the effect of transparent metamaterial absorber on the conversion efficiency of the solar array have also been measured. These results fully demonstrate the reliability of our design for solar arrays, which also meet the requirements of structural strength, atomic oxygen erosion resistance, weight limitation, etc.

## 1. Introduction

Electromagnetic absorber refers to a kind of material that can effectively absorb the incident electromagnetic wave[1,2]. The original microwave absorber is the Salisbury screen that appeared in World War II [3]. It consists of a thin resistive layer, a dielectric layer with a quarter-wavelength thickness, and a metal backplane. The incident electromagnetic wave can enter the interior of the absorbing material and ensure a very low reflectivity. The classical microwave absorber has been unable to meet the requirements of the modern environment because of its shortcomings, such as narrow bandwidth and large thickness [4].

In 2008, N. I. Landy of Boston University [5] proposed for the first time that adopting a sandwich structure to build an electromagnetic metamaterial unit can be used for high absorption of electromagnetic wave, while its thickness is only one of dozens of wavelengths, forming an ultra-thin absorber. This electromagnetic metamaterial absorber breaks the limitation that the traditional electromagnetic absorbing material needs 1/4 wavelength thickness at one stroke, and has the characteristics of high absorption efficiency and easy fabrication. However, it is very sensitive to the polarization of the incident wave and has a narrow absorption bandwidth. To overcome the shortcoming of metamaterial absorbers, an enhanced bandwidth metamaterial absorber is proposed in [6] by making two overlapped resonances. Furthermore, the researchers designed a series of electromagnetic metamaterials [7,8] which are insensitive to the polarization of incident wave, and their design of unit cell generally adopt a central symmetric structure to achieve the same high-efficiency absorption effect for TE and TM incident waves. Besides, some research groups have increased the absorption band of polarization insensitive electromagnetic metamaterial absorber to terahertz [9-10] and infrared band [11]. In general, several ways have been put into practice to achieve multiband or wideband absorption [12-16]: (1) combine multiple resonant structural units; (2) stack the multi-layer resonant structures; (3) load with lumped elements; (4) load with high resistance surface.

In recent years, as RCS reduction of the satellite is a research hotspot, the research on high transparency and low scattering metasurface has a higher practical value. The traditional low scattering metasurface mostly uses the way of loading high resistance surface or multi-layer complex structure such as lumped components to achieve the design goal. Furthermore, traditional low-scattering metasurface mainly uses opaque substrate, which brings problems of a complex structure, huge weight, and low optical transmittance. Inspired by the Salisbury screen, transparent conductive films [17] were first applied in optically transparent absorbers, while it failed to meet the practical requirements due to the one-quarter wavelength limit. With the development of metamaterial absorbers (MMAs), several researchers were enlightened by the electromagnetic properties of water to design a novel metamaterial microwave absorber with a composite water-based structure based on random distribution [18,19], which can achieve a variable absorption band. Furthermore, the ground metal plate was also replaced by the transparent conductive film, enabling the absorber to be used in window applications [20-23]. To realize lightweight, low profile and wideband, a series of MMAs are proposed [24,25]. Different patterned design to compose broadband metamaterial that can achieve perfect absorption and transmit light were proposed [26,27]. In addition, in the practical aspect, lightweight, thin, high light transmittance are also important factors. These studies fully confirm that the wideband absorber with good transmittance can be designed by using transparent conductive film and transparent medium material. However, the traditional sandwich absorber is still difficult to cover ultra-wideband absorption which can be realized by the multi-layer absorber. The only drawback is that the profile of multi-layer absorber is high, and what's worse, multi-layer ITO structure is not conducive to maintain good light transmittance.

In this paper, indium tin oxide (ITO) is proposed to replace the light-permeable material (acrylic plate, special glass, etc.) as the dielectric substrate and an optically transparent, wideband metamaterial absorber is presented. High optical transparency and broadband absorption are achieved at the same time. Also, our design can keep stable when exposed to the severe environment of outer space. It is expected that the absorber can be used in the RCS reduction of a satellite which is of great significance for preventing radar detection and surveillance.

## 2. Results

As shown in Figure 1(a), the visible light transparent structure is mainly composed of three layers. The top layer and bottom layer are 2 mm and 1.1 mm thick anti-reflection glass respectively. The middle of the two layers of glass is made of ITO (indium tin oxides) cell structure, of which the square resistance value is 25 Ω / square. As shown in Figure 1(b), the unit structure of the middle layer consists of a circular ITO structure and a pair of arc ITO structures with unequal lengths. The parameters of the unit structure are as follows: $a = b = 7.5$mm, $W_1 = 2.4$mm, $W_2 = 2.6$mm, $R_1 = R_2 = 3.7$mm, $R_3 = 0.8$mm, $α = 125°$, $β = 195°$. As the transparent structure is attached to the surface of the solar array, the solar array can be considered as a metal backplate. Therefore, the transparent microwave invisible structure we designed is a reflective absorber.

By full-wave electromagnetic simulation software CST Microwave Studio, we simulate the unit structure of the designed light transparent low scattering metasurface. When the horizontal (vertical) polarized wave is incident normally in the - Z direction, the simulated co-polarization and cross-polarization reflection coefficients are shown in Figure 2. As shown in Figure 2, we can see that from 7.0 GHz to 17.6 GHz, the values of the co-polarization reflection coefficient are all below -10dB, while the values of cross-polarization reflection coefficient are always below -10dB. In conclusion, when the plane wave with horizontal (or vertical) polarization is incident normally in the range of 7-18GHz, a small part of the incident wave undergoes polarization conversion and a large part of the incident wave has been absorbed. Since most of the transmitted energy is blocked by the ground, the formula of the absorption rate can be written as $A=1-|S_{11}|^2$. In Figure 2, the absorptivity of the designed metasurface is calculated and plotted. In the range of 7-18GHz, the absorptivity is more than 87%, and in the range of 11.2-17.4 GHz, the absorptivity can reach 90%.

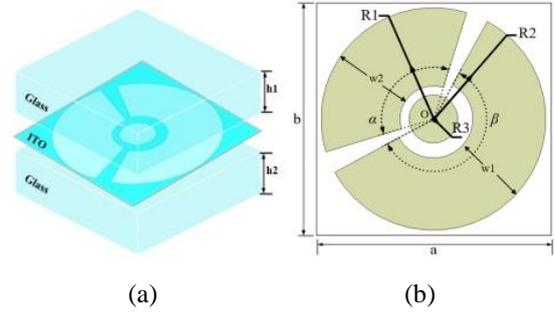

(a)                (b)

Fig 1: (a) 3D perspective view of the unit structure (b) structure parameters of the ITO layer.

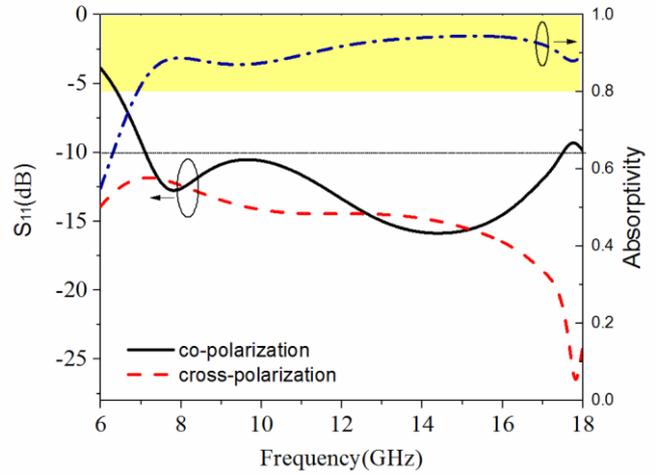

Fig 2: Simulated co- and cross-polarized $S_{11}$(left column) and absorptivity(right column).

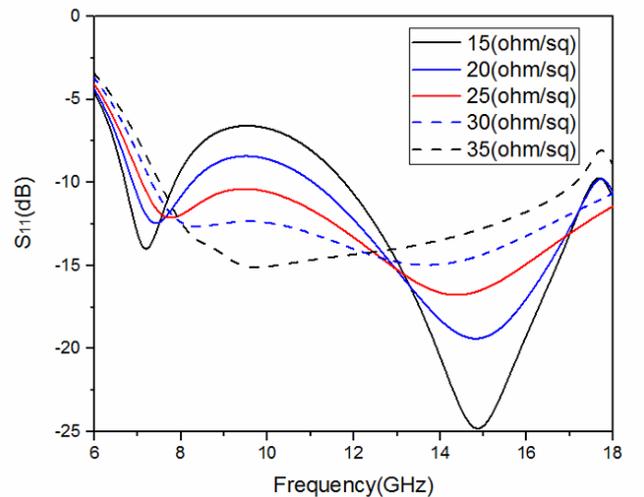

Fig 3: Simulated co-polarized reflection coefficients of the metamaterial absorbers with varied surface resistance of the ITO ($R_s$=15, 20, 25, 30 and 35 Ω/sq).

As the variety of surface resistance of the ITO, the absorption performance of the proposed transparent absorber is entirely different. Additionally, processing technology is not able to attain the surface resistance perfectly. Therefore, it is important to study the effect of varied resistance of ITO on $S_{11}$, to guide the design of absorber.

As can be seen from Figure 3, the surface resistance of the ITO film has big influence on the co-polarized reflection coefficients of the absorber, which means the co-polarized reflection coefficients are sensitive to the variation of surface resistance. So, it provides a limit for the error of manufacture.

To better understand the broadband absorption characteristic of the metamaterial absorber, a detailed simulation was performed to analyze the spatial distribution of the electric currents. At the two peak absorption frequencies, 8.0 GHz and 15.0 GHz, the surface current distributions at these resonant frequencies on the top metallic structure and metallic groundsheet are shown in Figure 4.

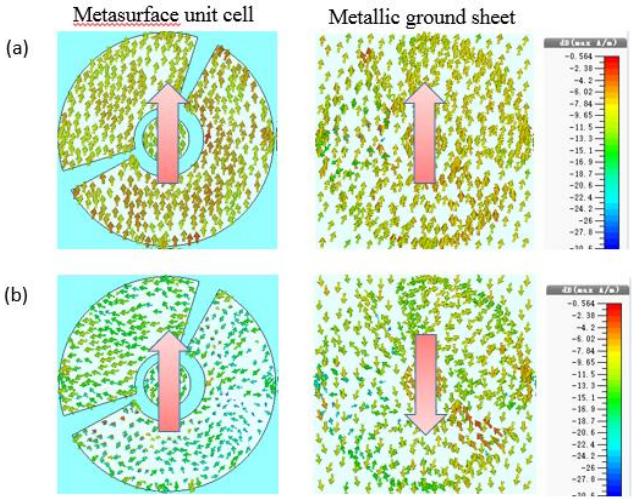

Fig 4: The current distributions on the metasurface (left column) and metallic ground sheet (right column) under y-polarized wave incidence when at (a) 8.0 GHz, (b) 15.0 GHz. The red arrow indicates the equivalent current direction.

As shown in Figure 4, the top metallic structure forms an equivalent surface current at 8.0 GHz, which is parallel to the current flowing on the bottom ground plane. Obviously, it can be regarded as an electric dipole that can excite the electric resonance. On the contrary, the surface current on the top metallic structure is antiparallel to the current on the ground plane at 15.0 GHz, in which case an equivalent circulating current is formed between the top metallic structure and the metallic ground, and magnetic resonances are created at 15.0 GHz.

Figure 5 shows the bi-static 3D scattering patterns of ITO absorber and the metal plate of the same size. As supplementary, simulated mono-static RCS of metal plate loaded with transparent metasurface is shown in Figure 6. In the frequency range of 11.2-18.2GHz, the mono-static RCS reduction effect can reach 10 dB or more.

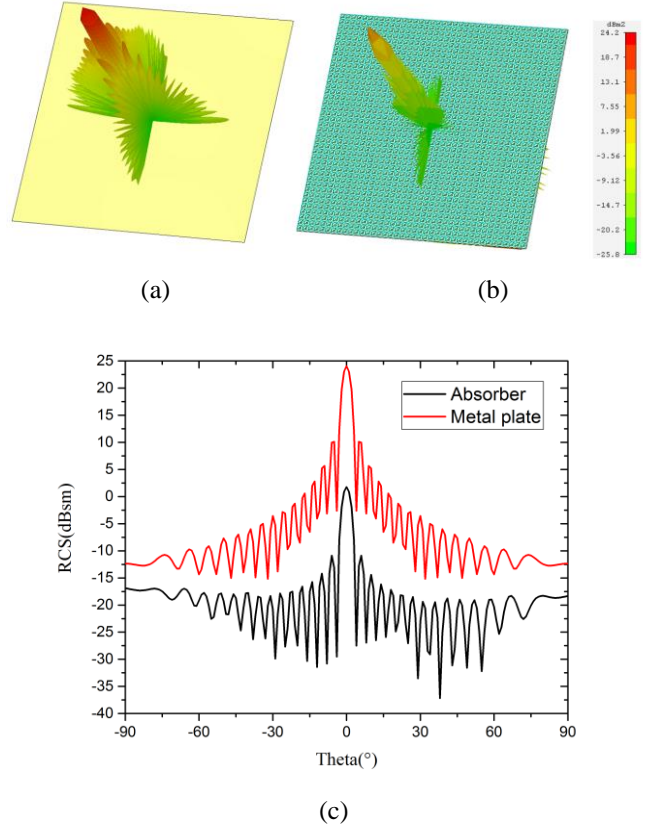

Fig 5: Simulated scattering patterns of PEC and transparent metamaterial absorber at 15.0 GHz for normal incidences. (a) Patterns of PEC, (b) Patterns of transparent metamaterial absorber, (c) Far-field cut of $\theta$ from -90° to 90°.

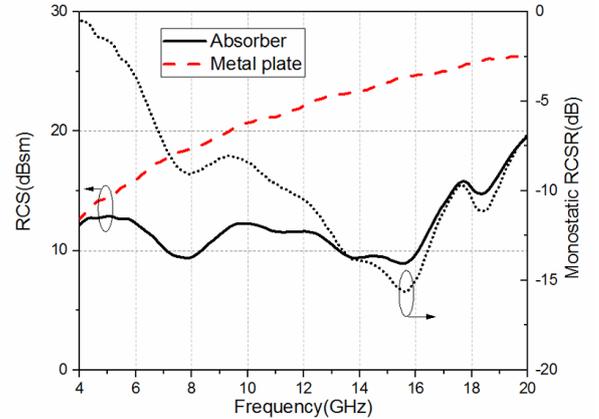

Fig 6: Simulated mono-static RCS of the proposed absorber and reference metal plate and effect of mono-static RCS reduction($\theta = \varphi = 0$).

## 3. Discussion

When the absorber is exposed to an incident electromagnetic wave, the transmission electromagnetic wave is totally blocked by the bottom solar array ground plane and the absorption rate of the absorber can be expressed as follows:

$$A(\omega) = 1 - R(\omega) - T(\omega) \qquad (1)$$

The above equation indicates that if we want a high-efficient absorption rate approaching unity, we should design the structure to suppress the reflection as much as possible. To achieve high microwave absorption, the equivalent impedance of the metamaterial absorber should be matched to the impedance of the air. When transmitted power and reflected power both equal to zero, the electromagnetic power is completely dissipated by the resistive thin film of the proposed absorber. When the solar array is coated with metamaterial absorber it can be regarded as a metal plate, so the transmitted power $T(w)$ is nearly zero.

Here, we establish a circuit model to give a physical insight into the absorption performance of the proposed metamaterial absorber, and therefore to form a semi-analytical theory for guiding the absorber design.

Both of the air and the dielectric layer can be regarded as a fraction of the transmission line with certain characteristic impedance, while the patterned ITO film in the middle layer can be modeled as a series circuit composed of inductance $L$, capacitance $C$ and resistance $R$. The shape of the patterned ITO film influences the values of the $L$ and $C$ parameters, while the resistance $R$ can be calculated by $R=Rs*S/A$ [28], where $Rs$, $S$ and $A$ represent the ITO sheet resistance (25 $\Omega/square$ in this scheme), area of the unit cell (7.5mm*7.5mm) and area of the ITO film, respectively.

Then, a general equivalent circuit model is established, as shown in Figure 7. It should be noted that we neglect the little loss of the glass on the top and bottom layer and assume that the unit cell is terminated by a shorten load.

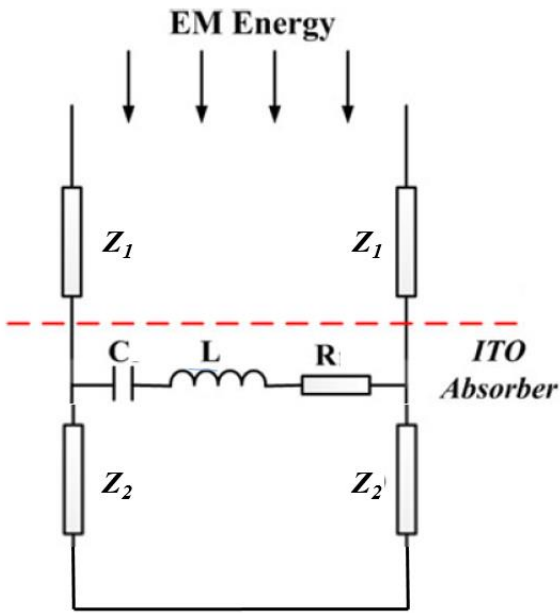

Fig 7: The equivalent circuit of the proposed ITO absorber.

TABLE 1
Calculated Parameters of the Equivalent Circuit

| R | L | C | $Z_1$ | $Z_2$ |
|---|---|---|---|---|
| 53Ω | 0.01nH | 0.35pF | 147Ω | 147Ω |

Quantitative simulation was performed by the Advanced Design System (ADS), and thus-obtained absorption curve was plotted and compared to the CST simulation result in Figure 8, where the red solid line is the curve obtained from ADS, and the black solid line is the curve obtained from CST. They match reasonably well.

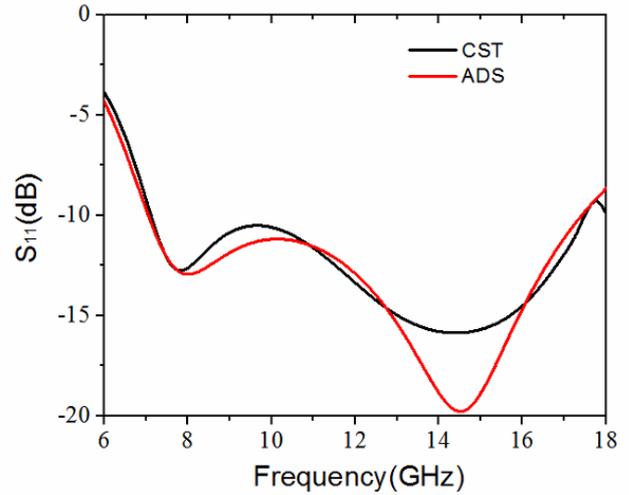

Fig 8: Comparison of the simulated $S_{11}$ by CST and equivalent circuit.

## 4. Materials and Methods

The design of the metasurface must realize broadband and low scattering while maintaining the high transmittance of the light-receiving surface of the solar array. To ensure that the transparent structure designed has good light permeability and the solar array can work normally, merely one ITO film is used in the fabrication.

Conductive indium-tin-oxide (ITO) film was prepared on the glass by magnetron sputtering. The square resistance of the ITO film was controlled by the time of etching. Then periodic ITO structure was etched on the glass by laser etching machine. At last, the sprayed ITO layer and bottom glass were manually stuck to top glass by ultrathin optically clear adhesive to ensure excellent light transmittance.

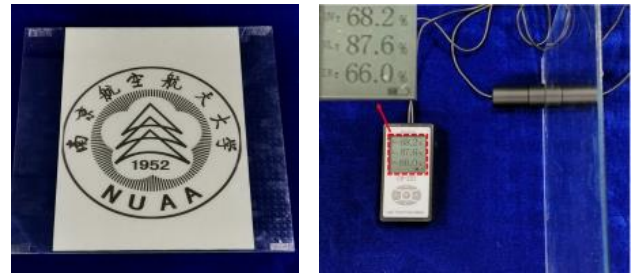

(a)　　　　　　　　　(b)

Fig 9: (a) Fabricated prototype of the proposed ITO absorber (b) Transmittance test.

As shown in Figure 9(a), the geometric dimension of the prototype is 300 × 300 mm, which is used along with the solar cell array sample. The ITO film is well-known for its transparency in the visible light region, and it has a strong reflection in the middle and far-infrared region, mainly due to the excitation of surface plasmonic waves. As shown in Figure 9(b), the optical transmittance is measured by the transmittance tester (LH-221), and the deviation of the instrument is less than 1%. Through random multi-point measurement, the measured transmittance of visible light is more than 80%, while that of infrared and ultraviolet light is less than 70%. In the visible light band, the average transmittance can reach 83.8%, and the highest transmittance can reach 87.6%.

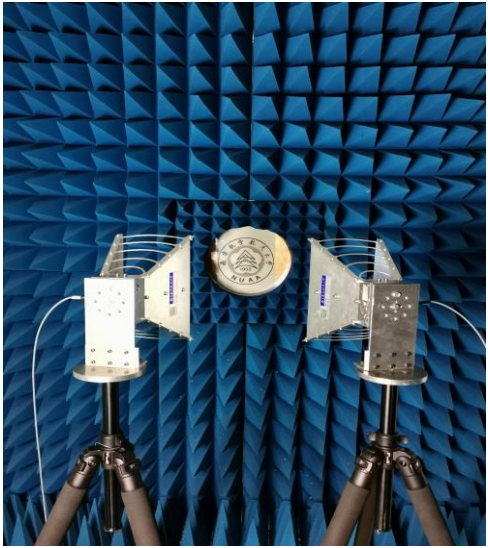

Fig 10: Measurement setup

The absorption experiment was performed in a microwave anechoic chamber, and the $S_{11}$ parameter was measured by a vector network analyzer (Agilent 5245A). Figure 10 shows the photo of the measurement setup. The sample is placed in the front of the horn antennas surrounding by absorbing materials, and the horn antennas are fixed on the two vertical arms of the trestle. Two horn antennas were used in the measurement to cover 1 GHz-20 GHz wavebands. The centers of the sample and horn antennas are set at the same height. One horn antenna is used as the source and the other is used for receiving reflected waves to obtain the co- and cross-polarization reflective coefficients. The absorption can be calculated by using Eq. (1). The measured absorption spectrum is shown in Figure 11, matching pretty well with the simulation results. Possible error origins include the noise during testing as well as the fabrication error.

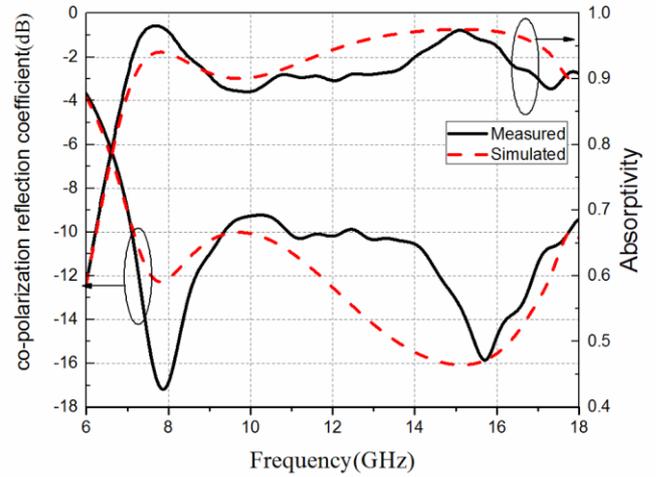

Fig 11: Test and simulation results of co-polarized $S_{11}$ (left) and absorptivity (right).

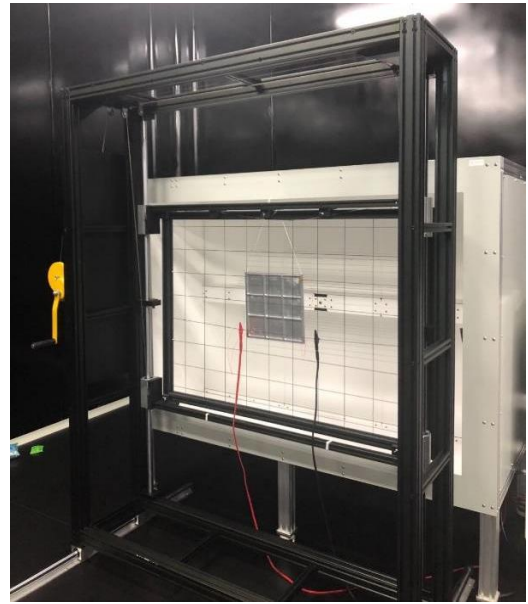

Figure 12: I-V test of the solar array

Measurement of the effect of transparent metamaterial absorber on the conversion efficiency of the solar array has also been performed. In order to understand the influence of the absorber on the photoelectric conversion efficiency, the I-V test of the solar array is needed. A comparison test was carried out before and after the solar cell array was covered with the transparent metamaterial absorber. Through the I-V curve obtained by the I-V test, we can obtain the influence of wide-band transparent metamaterial absorber on the photoelectric conversion efficiency of the solar array. The voltage and current points of the cell array can be drawn by adjusting the resistance value of the variable resistance, then the I-V curve of the cell array can be obtained, and the electrical performance parameters of the solar array can be obtained: open-circuit voltage *Voc*, short circuit current *Isc*, best working voltage *Vm*, best working current *Im* and maximum power *Pm*. The test result of the influence of the transparent metamaterial absorber on the conversion efficiency of the solar array can be seen in Figure 13.

By analyzing of I-V test results, after loading transparent low RCS absorber, the output power of the solar array decreased from 20.929 watts to 16.566 watts, 79.2% of the output power of the original solar array, which means the prototype has little effect on the conversion efficiency of the solar cell.

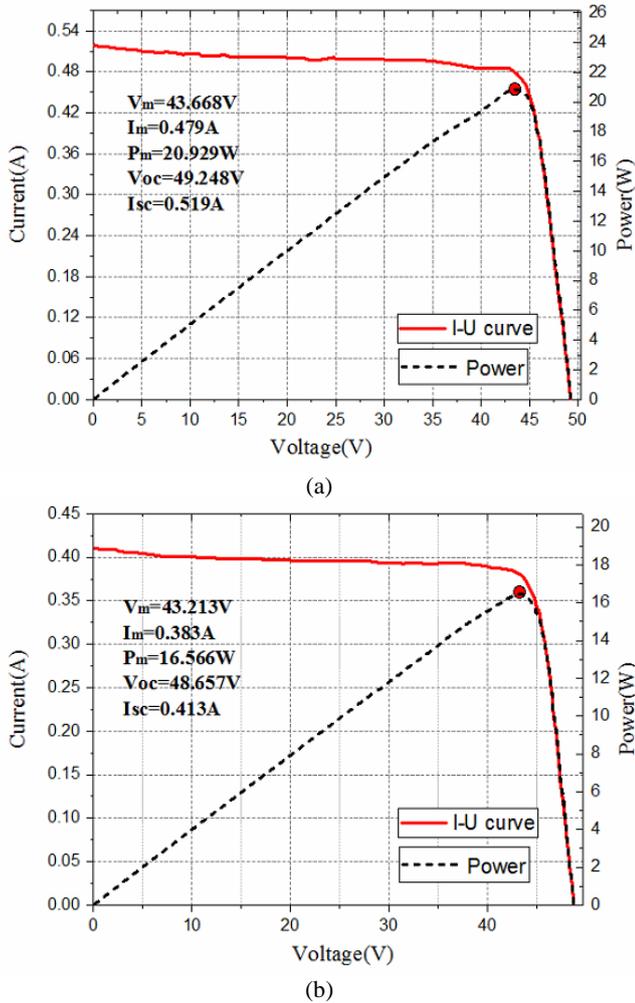

Fig 13: I-V test results of the solar array (a) without transparent absorber (b) with the transparent absorber.

## Conflicts of Interest

The authors declare that there is no conflict of interest regarding the publication of this article.

## Authors' Contributions

Xiangkun Kong and Shunliu Jiang contributed equally to this work. Xiangkun Kong, Qi Wang and Xiang Zhang designed the experiments. Shunliu Jiang, Lingqi Kong, and Haobin Hu performed the experiments. Shunliu Jiang analyzed the data and wrote the manuscript. Xiangkun Kong and Xin Zhao supervised the project.

## Acknowledgements

This work was supported by the Fundamental Research Funds for the Central Universities (No. NS2019023), National Natural Science Foundation of China (61901217) and by Open Research Program in China's State Key Laboratory of Millimeter Waves (Grant No. K202027).